# Methods for the generation of normalized citation impact scores in bibliometrics:

## Which method best reflects the judgements of experts?

Lutz Bornmann* & Werner Marx**

* Corresponding author:

Division for Science and Innovation Studies

Administrative Headquarters of the Max Planck Society

Hofgartenstr. 8,

80539 Munich, Germany.

Email: bornmann@gv.mpg.de

**Max Planck Institute for Solid State Research

Heisenbergstraβe 1,

70569 Stuttgart, Germany.

Email: w.marx@fkf.mpg.de

**Abstract**

Evaluative bibliometrics compares the citation impact of researchers, research groups and institutions with each other across time scales and disciplines. Both factors - discipline and period - have an influence on the citation count which is independent of the quality of the publication. Normalizing the citation impact of papers for these two factors started in the mid-1980s. Since then, a range of different methods have been presented for producing normalized citation impact scores. The current study uses a data set of over 50,000 records to test which of the methods so far presented correlate better with the assessment of papers by peers. The peer assessments come from F1000Prime - a post-publication peer review system of the biomedical literature. Of the normalized indicators, the current study involves not only cited-side indicators, such as the mean normalized citation score, but also citing-side indicators. As the results show, the correlations of the indicators with the peer assessments all turn out to be very similar. Since F1000 focuses on biomedicine, it is important that the results of this study are validated by other studies based on datasets from other disciplines or (ideally) based on multi-disciplinary datasets.

# 1      Introduction

Evaluative bibliometrics compares the citation impact of researchers, research groups and institutions with each other across timescales and disciplines. Both factors - discipline and period - have an influence on the citation count which is independent of the quality of the publications. Normalizing the citation impact of papers for these two factors started in the mid-1980s (Schubert & Braun, 1986). Since then, a range of different methods have been presented for producing normalized citation impact scores.

In this connection it is basically a matter of distinguishing two levels on which the normalization can be performed: (1) the level of the cited publication (cited-side). With this method, one counts the total citation count for the publication to be assessed (times cited) and then compares this value with those for similar publications (publications from the same subject area and publication year) - the reference set. (2) the level of the citing publication (citing-side). This method of normalization is oriented towards the citing and not the cited publication: Since the citations of a publication come from various subject areas, citing-side normalization aims to normalize each individual citation by subject and publication year.

As shown in section 2 below, a range of bibliometric methods for the normalization of the cited- and the citing-side have already been developed and presented. A bibliometrician who wants to use an advanced bibliometric indicator in a study is thus faced with the question of which approach to adopt. Each approach has particular methodological advantages and disadvantages which speak for or against its use. The comparison of metrics with peer evaluation has been widely acknowledged as a way of validating metrics (Garfield, 1979; Kreiman & Maunsell, 2011). Using data from F1000 – a post-publication peer review system of the biomedical literature – Bornmann and Leydesdorff (2013) investigated the relationship between ratings by peers and normalized impact scores against this background. The current study continues the line of this paper in that the validity of various methods of impact normalization is investigated with the help of ratings by peers from the F1000 post-publication peer review system. Compared with Bornmann and Leydesdorff (2013), this study uses a considerably larger data set, and also does not use cited-side alone, but also citing-side indicators. Besides the normalized indicators, we include observed citation counts (times cited) for comparison. The comparison is intended to show whether the normalized indicators measure more accurately research impact (as a proxy of quality) than an indicator without normalization (that means observed citation counts for a fixed citation window of three years).

## 2    Normalization of citation impact

Figure 1 shows the dependency of citation impact for papers on the subject category to which a Thomson Reuters journal is assigned (A), and the journal's publication year (B). The basis of these assessments is, for (A) all articles in the Web of Science (WoS, Thomson Reuters) from the year 2007, and for (B) all articles from the years 2000 to 2010. It is clearly visible from Figure 1 (A) that the average impact varies significantly with subject area. Whereas, for example, it is 10.77 for engineering and technology, for medical and health sciences it reaches 16.85. However, the citation impact is not only dependent on the subject category, but also on the publication year. As shown in Figure 1 (B), fewer citations may be expected, on average, for more recent publications. Whereas articles published in 2010 achieve a citation rate of only 7.34, articles from the year 2000 reach 22.53.

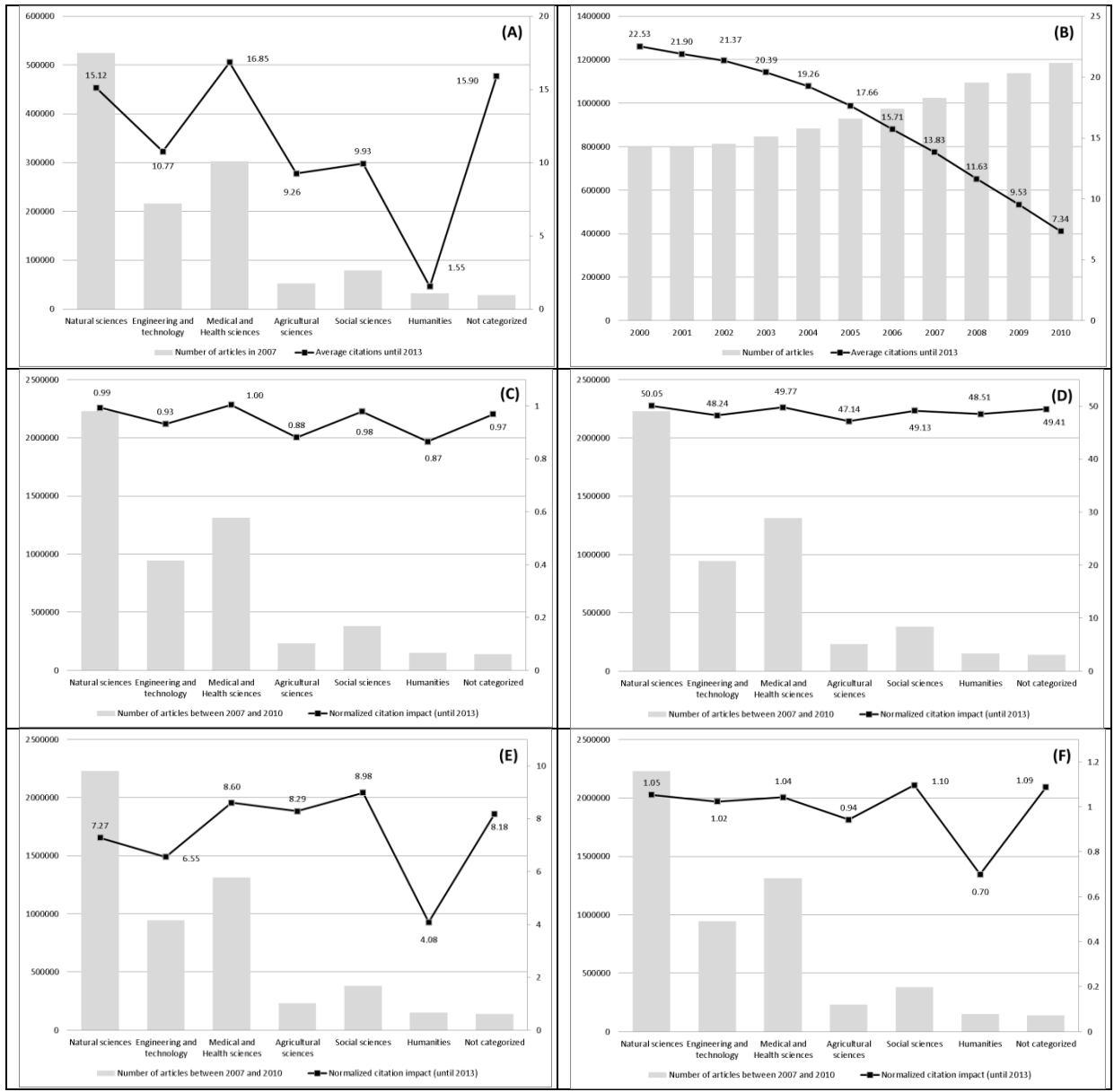

Figure 1. (A) Average citations of articles in different subject areas (and number of articles published). (B) Average citations of articles published between 2000 and 2010 (and number of articles published). (C) Average MNCS of articles (and number of articles published). (D) Average Hazen percentiles of articles (and number of articles published). (E) Average P100 of articles (and number of articles published). (F) Average SNCS3 of articles (and number of articles published).
Sources for the data: Web of Science (Thomson Reuters). The articles have been categorized into subject areas by using the OECD Category scheme which corresponds to the Revised Field of Science and Technology (FOS) Classification of the Frascati Manual (Organisation for Economic Co-operation and Development, 2007).

Since it is not only this study which has found different citation rates for different subject categories and publication years, but also nearly all the other studies which have appeared so far, these are the factors which are generally used for the normalization of citation impact. We can distinguish between two fundamental approaches for normalization: With cited-side normalization, the normalization is performed on the basis of the cited papers, and with citing-side on the basis of the citing papers. In the context of each type of normalization, different indicators are suggested, the most important of which are included in this study. The indicators are introduced in the following.

## 2.1 Cited-side normalization of citation impact

Cited-side normalization generally only takes account of citable documents (such as articles, reviews, and letters). Fundamentally, cited-side normalization compares the citation impact of a focal paper with an expected citation impact value. The expected value is the average citation impact of the papers in the same subject category as the paper in question and which appeared in the same publication year. This set of papers is referred to as the reference set. The calculation of a quotient of observed and expected citations represents the current bibliometric standard for performing the normalization of citation impact. A quotient of 1 corresponds to an average citation impact of the papers in the same subject area and publication year. A quotient of 1.5 indicates that the citation impact is 50% above the average (Waltman, van Eck, van Leeuwen, Visser, & van Raan, 2011). This quotient is used both in the Leiden Ranking (Waltman et al., 2012), and in the SCImago Institutions Ranking (SCImago Reseach Group, 2013), under the designations Mean Normalized Citation Score (MNCS, Leiden Ranking) and Normalized Impact (NI, SCImago Institutions Ranking) (Bornmann, de Moya Anegón, & Leydesdorff, 2012). In what follows, the abbreviation MNCS is used for this indicator.

Figure 1 (C) shows the MNCS of articles published between 2007 and 2010 sorted by subject category. Although the figure shows the OECD category scheme, the WoS journal subject categories have been used to calculate the MNCS (these categories have been also used for the calculation of the other indicators with cited-side normalization which will be discussed below). As expected, the MNCS values are close to 1 in all subject categories (they range from 0.87 to 1). This result indicates that cited-side normalization with the MNCS can perform a normalization of the citation impact both in respect of time as well as discipline.

The distribution of citation data is generally extremely skewed: most papers are hardly or not at all cited, whereas a few papers are highly cited (Seglen, 1992). Since the arithmetic mean is not appropriate as a measure of the central tendency of skewed data, percentiles of citations have been suggested as an alternative to MNCS (which is based on the arithmetic mean values of citations). The percentile indicates the share of the papers in the reference set which have received fewer citations than a paper in question. For example, a percentile of 90 means that 90% of the papers in the reference set have received fewer citations than the paper in question. The citation impacts of papers which have been normalized using percentiles are directly comparable with one another. For example, if two papers have been normalized with different reference sets and have a percentile of citations of 70, both have - compared with the other papers in the reference set - achieved the same citation impact. Even though both papers may have different citation counts, the citation impacts are the same.

Percentiles may be calculated with various procedures (Bornmann, Leydesdorff, & Mutz, 2013). For the current study, two procedures were used which may be described as the most important. For both procedures, the rank-frequency function is first calculated. All publications in the reference set are ranked in decreasing or increasing order by their number of citations ($i$), and the number of publications in the reference set is determined ($n$). For the product InCites (a customized, web-based research evaluation tool based on bibliometric data from WoS), Thomson Reuters generates the percentiles by using (basically) the formula ($i/n * 100$) (described as "InCites" percentiles in the following). Since, however, the use of this formula leads to the mean percentile of a reference set not being 50, the formula (($i - 0.5)/n * 100$) derived by Hazen (1914), which does not suffer this disadvantage, is used for calculating percentiles. The abbreviation "Hazen" is used for these percentiles in the following. Since the papers are sorted in increasing order of impact for the InCites percentiles, and in decreasing order for Hazen percentiles, the InCites percentiles are inverted, subtracting the values from 100. An exact presentation of the calculation of these and other percentiles in bibliometrics can be found in Bornmann, Leydesdorff, and Mutz (2013).

Figure 1 (D) shows average Hazen percentiles of citations for various disciplines. The underlying data set includes all articles in the WoS from the years 2007 to 2010. All disciplines have an average percentile of around 50. The normalized citation impact, which indicates an average citation impact, is thus the same for all disciplines. So normalization has achieved the desired effect.

Bornmann, Leydesdorff, and Wang (2013) introduced P100 as a new citation-rank approach. One important advantage of P100 compared with other normalized indicators is that the scale values in a reference set are distributed from 0 to 100 exactly and are thus comparable across different reference sets. The paper with the highest impact (lowest impact) in one reference set receives the same scale value as the paper with the highest impact (lowest impact) in another reference set. With the InCites and Hazen percentiles, the most and the least cited papers in a reference set generally receive very different values. For the P100 indicator, citations of papers in a reference set are ranked according to their frequencies of papers, which results in a size-frequency distribution (Egghe, 2005). This distribution is used to generate a citation rank where the frequency information is ignored. In other words, instances of papers with the same citation counts are not considered. This perspective on citation impact focuses on the distribution of the unique citation counts with the information of maximum, median, and minimum impact and not on the distribution of the papers (having the same or different citation impact) which is the focus of interest in the conventional citation analysis.

To generate citation ranks for a reference set, the unique citations are ranked in ascending order from low to high citation counts and ranks are attributed to each citation count, with rank 0 for the paper with the lowest impact or zero citations. In order to generate values on a 100-point scale (P100), each rank i is divided by the highest rank $i_{max}$ and multiplied by 100, i.e. ($100*(i/i_{max})$).

Figure 1 (E) shows average P100s of articles which were published in different subject categories and publication years. Even if for some disciplines, such as medical and health sciences, agricultural sciences and social sciences, P100≈8 yields a similar average value, P100≈4 yields a substantial deviation from this value with the humanities. Thus it is clear that the normalization of citation impact is not successful in all disciplines. As Bornmann and Mutz (in press) and also Schreiber (2014) were able to show, P100 has some weaknesses, including the paradoxical situation that the scale value of a paper can increase as the result of another paper receiving an additional citation. Bornmann and Mutz (in press) therefore suggest the indicator P100' as an improvement on P100. In contrast to P100, the ranks for P100' are not only based on the unique citation distribution, but also consider the frequency of papers with the same citation counts. For P100', each rank i is divided by the highest rank ($i_{max}$ or (n-1)) papers in the reference set and is multiplied by 100, i.e. $100*(i/i_{max})$. According to the evaluations of Schreiber (in press), however, P100' (unlike P100) strongly resembles the percentile-based indicators (such as Hazen and InCites).

## 2.2 Citing-side normalization of citation impact – the weighting of individual citations

Even if the current methods of cited-side normalization differ in their calculation of normalized citation impact, they are still derived from the same principle: for a cited paper whose citation impact is of interest, a set of comparable papers is compiled (from the same subject category and the same publication year). By contrasting the observed and the expected citations, cited-side normalization attempts to normalize the citation impact of papers for the variations in citation behaviour between fields and publication years. However, this does not take into account that the citation behaviour is different on the level of the citing papers. In most cases, the citations for a paper do not come from one field, but from a number of fields. Thus, for example, the paper of Hirsch (2005), in which he suggests the h index for the first time, is cited from a total of 27 different subject areas (see Table 1). In other words, the citations originate in quite different citation cultures.

Table 1. Subject areas of the journals in which the papers citing Hirsch (2005) have appeared. The search was performed on 14.8.2013 in Scopus (Elsevier). Since the journals of the 1589 citing papers were assigned to an average of 1.8 subject areas, the result was a total of 2778 assignments.

| Subject area | Number of citing papers |
| --- | --- |
| Computer Science | 698 |
| Social Sciences | 506 |
| Medicine | 338 |
| Mathematics | 229 |
| Decision Sciences | 191 |
| Biochemistry, Genetics and Molecular Biology | 103 |
| Agricultural and Biological Sciences | 97 |
| Engineering | 85 |
| Business, Management and Accounting | 63 |
| Environmental Science | 61 |
| Psychology | 51 |
| Physics and Astronomy | 49 |
| Multidisciplinary | 42 |
| Economics, Econometrics and Finance | 38 |
| Arts and Humanities | 31 |
| Chemistry | 30 |
| Earth and Planetary Sciences | 28 |
| Nursing | 24 |
| Health Professions | 23 |
| Pharmacology, Toxicology and Pharmaceutics | 20 |
| Materials Science | 18 |
| Chemical Engineering | 16 |
| Neuroscience | 13 |
| Immunology and Microbiology | 13 |
| Energy | 4 |
| Dentistry | 4 |
| Veterinary | 3 |
| Total | 2778 |

As Figure 1 (A) shows, citations are more probable in the disciplines medical and health sciences and natural sciences than in the social sciences and humanities. The evaluations of Marx and Bornmann (in press) indicate that citing is no less frequent in these disciplines than in other disciplines, but that the share of cited references covered in WoS is especially low. In this case "covered" means that the cited reference refers to a journal which is evaluated by Thomson Reuters for the WoS. Measured by the total references available, the social sciences, for example, exhibit the highest cited reference rate of all the disciplines considered. Not only in the social sciences, but also in the agricultural sciences and the humanities, many references point to document types other than papers from the journals covered in the WoS, such as books and book chapters (which are not generally captured by WoS as database documents), as well as journals which do not belong to the evaluated core journals of the WoS.

Given the different expected values for citation rates in different disciplines, the <u>citations</u> should be normalized accordingly, in order to obtain a comparable citation impact between different citing papers. The idea of normalizing citation impact on the citing-side stems from a paper by Zitt and Small (2008), in which a modification of the Journal Impact Factor (Thomson Reuters) by fractional citation weighting was proposed. Citing-side normalization is also known as fractional citation weighting, source normalization, fractional counting of citations or a priori normalization (Waltman & van Eck, 2013a). It is not only used for journals (see Zitt & Small, 2008), but also for other publication sets. This method takes into account the citation environment of a citation (Leydesdorff & Bornmann, 2011; Leydesdorff, Radicchi, Bornmann, Castellano, & de Nooy, in press), by giving the citation a weighting which depends on its citation environment: A citation from a field in which citation is frequent receives a lower weighting than a citation from a field where citation is less common.

In the methods proposed so far for citing-side normalization, the number of references of the citing paper is often used as a weighting factor for the citation (Waltman & van Eck, 2013b). Here the assumption is made that this number for a paper reflects the typical number in the field. Since this assumption cannot always be made, the average number of references from other papers which appear in a journal alongside the citing paper is also used as a weighting factor. This approach has a high probability of improving the accuracy of estimation of the typical citation behaviour in a field (Bornmann & Marx, in press). In the following, three variants of a method of citing-side normalization are presented, which were suggested by Waltman and van Eck (2013b). These variants are included in the current study.

**Variant 1:**

$$SNCS1 = \sum_{i=1}^{c} \frac{1}{a_i}$$

With the SNCS1 (Source Normalized Citation Score) indicator, $a_i$ is the average number of linked references in those publications which appeared in the same journal and in the same publication year as the citing publication i. Linked references refer to papers from journals which are covered by the WoS. The limitation to linked references (instead of all references) should prevent the disadvantaging of fields which often cite publications which are not indexed in WoS. As the evaluations of Marx and Bornmann (in press) have shown, this danger of disadvantaging really does exist (see above): thus, for example, in the social sciences the average number of linked cited references is significantly lower than the average overall number of cited references.

To calculate the average number of linked references in SNCS1, not all are used, but only those from particular reference publication years. The number of the reference publication years orients themselves towards the number of those years which are determined for the citations of a publication. For example, if the citation window for a publication (from 2008) covers a period of four years (2008 to 2011), then every citation of this publication (e.g. a citation from 2010) is divided by the average number of linked references to the four previous years (in this case 2007 to 2010). The limitation to the recent publication years is intended to prevent fields in which older literature plays a large role from being disadvantaged in the normalization (Waltman & van Eck, 2013b).

**Variant 2:**

$$SNCS2 = \sum_{i=1}^{c} \frac{1}{r_i}$$

With SNCS2, each citation of a publication is divided by the number of linked references in the citing publication (instead of by the number of linked references of all publications <u>of the journal in question</u> as in the case of SNCS1). The selection of the reference publication years is, analogously to SNCS1, oriented towards the size of the citation window.

**Variant 3:**

$$SNCS3 = \sum_{i=1}^{c} \frac{1}{p_i r_i}$$

SNCS3 can be seen as a combination of SNCS1 and SNCS2. $r_i$ is defined analogously to SNCS2. $p_i$ is the share of the publications which contain at least one linked reference among those publications which appeared in the same journal and in the same publication year as the citing publication i. The selection of the reference publication years is, analogously to SNCS1 and SNC2, oriented towards the size of the citation window.

According to the empirical results of Waltman and van Eck (2013b) and Waltman and van Eck (2013a), citing-side normalization has shown more successful than cited-side normalization,

Whereas Waltman and van Eck (2013b) only included selected core journals in the WoS database for the calculation of the SNCS indicators, the indicators for the present study were calculated on the basis of all the journals in the WoS database. As the SNCS3 scores for all articles in the WoS from 2007 to 2010 in Figure 1 (F) show, the average scores for SNCS3≈1 are similar for all disciplines. So it seems that the normalization method basically works. However, as with the P100 indicator, here too the results for the humanities are different.

# 3 Methods

## 3.1 Peer ratings provided by F1000

F1000 is a post-publication peer review system of the biomedical literature (papers from medical and biological journals). This service is part of the Science Navigation Group, a group of independent companies that publish and develop information services for the professional biomedical community and the consumer market. F1000 Biology was launched in 2002 and F1000 Medicine in 2006. The two services were merged in 2009 and today constitute the F1000 database. Papers for F1000 are selected by a peer-nominated global "Faculty" of leading scientists and clinicians who then rate them and explain their importance (F1000, 2012). This means that only a restricted set of papers from the medical and biological journals covered is reviewed, and most of the papers are actually not (Kreiman & Maunsell, 2011; Wouters & Costas, 2012).

The Faculty nowadays numbers more than 5,000 experts worldwide, assisted by 5,000 associates, organized into more than 40 subjects (which are further subdivided into over 300 sections). On average, 1,500 new recommendations are contributed by the Faculty each month (F1000, 2012). Faculty members can choose and evaluate any paper that interests them. Although many papers published in popular and high-profile journals (e.g. *Nature*, *New England Journal of Medicine*, *Science*) are evaluated, 85% of the papers selected come from specialized or less well-known journals (Wouters & Costas, 2012). "Less than 18 months since Faculty of 1000 was launched, the reaction from scientists has been such that two-thirds of top institutions worldwide already subscribe, and it was the recipient of the Association of Learned and Professional Society Publishers (ALPSP) award for Publishing Innovation in 2002 (http://www.alpsp.org/about.htm)" (Wets, Weedon, & Velterop, 2003, p. 249).

The papers selected for F1000 are rated by the members as "Good," "Very good" or "Exceptional" which is equivalent to scores of 1, 2, or 3, respectively. In many cases a paper is assessed not just by one member but by several. Overall the F1000 database is regarded simply as an aid for scientists to receive pointers to the most relevant papers in their subject area, but also as an important tool for research evaluation purposes. So, for example, Wouters and Costas (2012) write that "the data and indicators provided by F1000 are without doubt rich and valuable, and the tool has a strong potential for research evaluation, being in fact a good complement to alternative metrics for research assessments at different levels (papers, individuals, journals, etc.)" (p. 14).

**3.2    Formation of the data set to which bibliometric data and altmetrics are attached**

In January 2014, F1000 provided one of the authors with data on all recommendations made and the bibliographic information for the corresponding papers in their system (n=149,227 records). The data set contains a total of 104,633 different DOIs, among which all are individual papers with very few exceptions. The approximately 30% reduction of the data set with the identification of unique DOIs can mainly be attributed to the fact that many papers received recommendations from several members and therefore appear multiply in the data set.

For bibliometric analysis in the current study, the normalized indicators (with a citation window between publication and the end of 2013) and the citation counts for a three years citation window were sought for every paper in an in-house database of the Max Planck Society (MPG) based on the WoS and administered by the Max Planck Digital Library (MPDL). In order to be able to create a link between the individual papers and the bibliometric data, two procedures were selected in this study: (1) A total of 90,436 papers in the data set could be matched with one paper in the in-house database using the DOI. (2) With 4,205 papers of the total of 14,197 remaining papers, although no match could be achieved with the DOI, one could be with name of the first author, the journal, the volume and the issue. Thus bibliometric data was available for 94,641 papers of the 104,633 total (91%). This percentage approximately agrees with the value of 93% named by Waltman and Costas (2014), who used a similar procedure to match data from F1000 with the bibliometric data in their own in-house database.

The matched F1000 Data (n=121,893 records on the level of individual recommendations from the members) refer to the period 1980 to 2013. Since the citation scores which were normalized on the citing-side are only available for the years 2007 to 2010 in the in-house database, the data set is reduced to n=50,082 records.

### 3.3  Statistical procedures and software used

The statistical software package Stata 13.1 (http://www.stata.com/) is used for this study; in particular, the Stata commands ci2, regress, margins, and coefplot are used. To investigate the connection between members' recommendations and normalized indicators, two analyses are undertaken:

(1) The Spearman's rank correlation coefficient with 95% confidence interval is calculated for the connection between members' recommendations and each indicator. The Pearson product-moment correlation coefficient is inappropriate for this analysis since neither the recommendations nor the indicators follow a normal distribution (Sheskin, 2007).

(2) A series of regression models have been estimated, to investigate the relationship between the indicators and the members' recommendations. For each indicator a regression model was calculated here. In order to be able to compare the results from models based on different indicators, the indicator scores are subjected to a z-transformation. The z-scores are rescaled values to have a mean of zero and a standard deviation of one. Each z-score indicates its difference from the mean of the original variable in number of standard deviations (of the original variable). A value of 0.5 indicates that the value from the original variable is half a standard deviation above the mean. To generate the z-scores, the mean is subtracted from the value for each paper, resulting in a mean of zero. Then, the difference between the individual's score and the mean is divided by the standard deviation, which results in a standard deviation of one.

The violation of the assumption of independent observations by including several F1000 recommendation scores associated with a paper is considered in the regression models by using the cluster option in Stata (StataCorp., 2013). This option specifies that the recommendations are independent across papers but are not necessarily independent within the same paper (Hosmer & Lemeshow, 2000, section 8.3). Since the z-transformed indicator violates the normality assumption, bootstrap estimations of the standard errors have been used. Here several random samples are drawn with replacement (here: 100) from the data set.

In this study, predictions of the previously fitted regression models are used to make the results easy to understand and interpret. Such predictions are referred to as margins, predictive margins, or adjusted predictions (Bornmann & Williams, 2013; Williams, 2012; Williams & Bornmann, 2014). The predictions allow a determination of the meaning of the empirical results which goes beyond the statistical significance test. Whereas the regression models illustrate which effects are statistically significant and what the direction of the effects is, predictive margins can provide a practical feel for the substantive significance of the findings. The predictive margins will be presented graphically.

## 4 Results

### 4.1 Mean citation rates

In a first step of analysis, we have compared the mean citation rates of the subject categories or subject category combinations, respectively, to which the journals of the F1000 papers have been assigned (by Thomson Reuters). Subject category combinations occur when journals have more than one category. Since the F1000 papers are generally published in the biomedical area, one could expect similar mean citation rates (and could question the usefulness of the dataset for the evaluation of normalization techniques). Table 2 shows mean citation rates, minimum and maximum number of citations for F1000 papers in different subject categories or subject category combinations, respectively. Of the total of 627 subject categories or subject category combinations, respectively, the 20 categories with the most papers are presented. As the results in the table shows, the differences in the mean citation rates are large: Whereas the papers in anaesthesiology reach a mean citation rate of 14.69, this rate is 107.22 in medicine, general & internal. Thus, the dataset seems to be appropriate to analyse normalization techniques – at least normalization techniques on the cited-side.

Table 2. Mean citation rates, minimum and maximum number of citations (for a three year citation window) for F1000 papers in different subject categories or subject category combinations, respectively. The 20 categories (or category combinations) are presented with the most F1000 papers (ordered by the number of papers).

| Subject category or subject category combination | Mean citation rate | Minimum | Maximum | Number of papers |
|---|---|---|---|---|
| Multidisciplinary sciences | 72.97 | 0 | 2,113 | 5,946 |

| Biochemistry & molecular biology, Cell biology | 58.76 | 0 | 1,765 | 2,005 |
|---|---|---|---|---|
| Neurosciences | 36.10 | 0 | 460 | 1,902 |
| Biochemistry & molecular biology | 24.41 | 0 | 411 | 1,711 |
| Cell biology | 38.91 | 0 | 464 | 1,097 |
| Urology & nephrology | 22.35 | 0 | 217 | 1,097 |
| Oncology | 54.34 | 0 | 664 | 991 |
| Immunology | 56.40 | 0 | 436 | 887 |
| Genetics & heredity | 70.00 | 0 | 1358 | 838 |
| Endocrinology & metabolism | 25.43 | 1 | 247 | 825 |
| Gastroenterology & hepatology | 31.69 | 0 | 425 | 769 |
| Anaesthesiology | 14.69 | 0 | 159 | 646 |
| Medicine, general & internal | 107.22 | 0 | 1,360 | 622 |
| Hematology | 34.25 | 0 | 240 | 610 |
| Chemistry, multidisciplinary | 26.61 | 0 | 165 | 561 |
| Dermatology | 15.60 | 0 | 209 | 503 |
| Cardiac & cardiovascular systems | 38.27 | 0 | 417 | 476 |
| Immunology, Medicine, research & experimental | 49.90 | 2 | 377 | 436 |
| Microbiology | 26.72 | 0 | 418 | 433 |
| Clinical neurology | 33.73 | 0 | 343 | 396 |

## 4.2    Correlation

Table 3 shows the Spearman's rank correlation coefficients for the relationship between the F1000 members' recommendations and the individual standardised indicators. Since a series of papers are often represented multiply in the data set with recommendations from different members, the results are given both for all recommendations, as well as just for the first recommendation of a paper. A comparison of the results allows the influence of multiple publications to be estimated.

Table 3. Spearman's rank correlation coefficients with 95% confidence intervals for the relationship between the members' recommendations and the individual standardised indicators

| Indicator | Coefficient for all recommendations of a paper (n=50,082) | Coefficient for the first recommendation of a paper (n=39,200) |
|---|---|---|
| Citations | .300 [.292, .308] | .245 [.236, .254] |
| InCites | .231 [.222, .239] | .192 [.183, .202] |
| Hazen | .229 [.221, .238] | .191 [.181, .200] |
| MNCS | .238 [.230, .246] | .194 [.185, .204] |
| P100 | .225 [.216, .233] | .183 [.173, .192] |
| P100' | .231 [.222, .239] | .192 [.182, .201] |
| SNCS1 | .269 [.261, .277] | .218 [.208, .227] |
| SNCS2 | .274 [.265, .282] | .221 [.211, .230] |
| SNCS3 | .266 [.258, .274] | .214 [.205, .224] |

As the results in the table show, the coefficients for all indicators are reduced when only the first recommendation is taken into account. Since we can expect more similar recommendations for the same paper than for different papers (many papers have received scores from more than one F1000 members), the reduction in which all indicators appear to a similar extent is easily explained. According to the guidelines which Cohen (1988) has published for the interpretation of correlation coefficients, the coefficients fall in an area between small (r=.1) and medium (r=.3). Although the citation indicator shows the largest correlation with the recommendation scores, the differences in coefficient height between the indicators are slight (within the two groups of recommendations).

## 4.3    Regression model

The calculation of the correlation coefficients between the recommendations and the indicators provides the first impression of the particular relationships. However, this evaluation does not make it clear how strongly the indicator scores differ between the papers assessed by the F1000 members as good, very good or excellent. In order to reveal these differences, nine regression models were calculated, each with one indicator (z-transformed) as the dependent and the members' recommendations as the independent variable. The results of the models are shown in Table 4.

Table 4. Results (coefficients) from nine regression models with one indicator (z-transformed) as the dependent and the members' recommendations as the independent variable (n=50,082)

| | (1) | (2) | (3) | (4) | (5) | (6) | (7) | (8) | (9) |
|---|---|---|---|---|---|---|---|---|---|
| | Citations | InCites | Hazen | MNCS | P100 | P100' | SNCS1 | SNCS2 | SNCS3 |
| | | | | | | | | | |
| Recommendation | | | | | | | | | |
| Good (reference category) |  |  |  |  |  |  |  |  |  |
|  |  |  |  |  |  |  |  |  |  |
| Very good | 0.33*** | 0.36*** | 0.36*** | 0.27*** | 0.35*** | 0.36*** | 0.29*** | 0.31*** | 0.29*** |
|  | (24.22) | (39.36) | (35.25) | (20.28) | (33.20) | (34.14) | (17.55) | (23.71) | (22.09) |
|  |  |  |  |  |  |  |  |  |  |
| Excellent | 0.87*** | 0.60*** | 0.60*** | 0.62*** | 0.72*** | 0.60*** | 0.76*** | 0.81*** | 0.75*** |
|  | (15.82) | (45.49) | (37.84) | (10.27) | (29.11) | (40.97) | (14.35) | (16.25) | (16.50) |
|  |  |  |  |  |  |  |  |  |  |
| Constant | -0.17*** | -0.16*** | -0.16*** | -0.13*** | -0.17*** | -0.16*** | -0.15*** | -0.16*** | -0.15*** |
|  | (-34.56) | (-26.37) | (-23.31) | (-27.56) | (-27.81) | (-21.89) | (-87.44) | (-34.13) | (-30.80) |

Notes: $t$ statistics in parentheses
*** $p < 0.001$

In order to visualise the differences between the indicator scores, after the regression analyses predictive margins were calculated, which can be seen in Figure 2. Due to the z-transformations of the indicators, the scores (predictive margins) for the different indicators are directly comparable with one another. The scores are displayed in the figure with 95% confidence intervals. These confidence intervals express something about the accuracy of the scores for an indicator. Whereas the confidence intervals of the indicators within a recommendation category (e.g. "good") may be compared with one another (because of the common number of records), this is not possible for confidence intervals across recommendations: with a better evaluation, greater confidence intervals are to be expected, since the number of records will be lower (good=29,515, very good=17,329, and excellent=3,238).

As the results in Figure 2 show, the predictive margins for the recommendation "good" are in relatively good agreement between the indicators with a value of around -1.6 which indicates that the value from the original normalized score is around one and a half standard deviations below the mean. Thus the indicators are in good agreement about the later impact of papers which are evaluated by the members as "good".

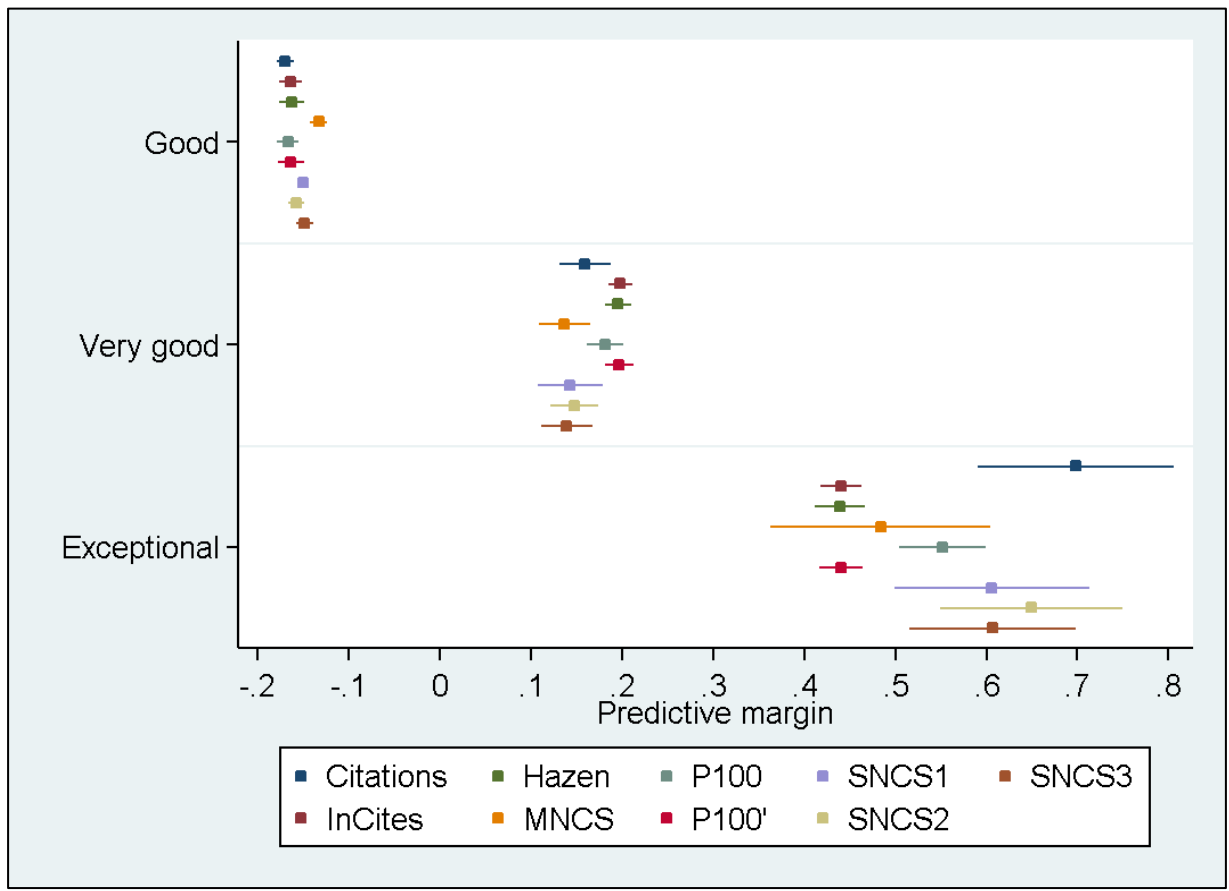

Figure 2. Predictive margins with 95% confidence intervals from the regression models.

The image of relatively good agreement between the indicators changes in regard to the papers evaluated as "very good". On the one hand, the percentile and the P100 based indicators are somewhat further removed from the mean value (0) than the MNCS or the SNCS indicators. On the other hand, some indicators (like SNCS3) exhibit a smaller accuracy than other indicators (such as P100'). The differences between the indicators increase still further – as Figure 2 shows – with the papers evaluated as "exceptional". The greatest deviation from the mean value appears with the SNCS and citation indicators. Apparently these indicators can differentiate better between "exceptional" and lower classified papers than the other indicators. Especially, the difference between the predictive margins for observed citations on the one hand and a number of cited-side normalized indicators (InCites, Hazen, and P100') on the other hand is rather large (0.70 vs. 0.45). However, for the SNCS and citation indicators the confidence intervals are relatively wide, which indicates a relatively small accuracy of the values.

## 5      Discussion

Bibliometrics on a professional level does not only evaluate the observed citations from publications, but also calculates normalized indicators which take into account that citations have different expected values depending on subject area and publication year (Council of Canadian Academies, 2012). For example, in the Snowball Metrics Recipe Book – a collection of recommendations for indicators which may be used for institutional evaluations (especially in the UK) – the use of a field-weighted citation impact score is recommended (Colledge, 2014). Up to now it has been customary to use the MNCS as a standardised indicator in evaluations. However, in recent years a range of alternatives to the MNCS have been presented, in attempts to avoid particular weaknesses of the indicator. Thus, for example, an extremely highly cited publication can influence MNCS so strongly that the score can hardly represent the totality of the publications of a set (Waltman, et al., 2012).

How far a standardised indicator other than the MNCS, such as Hazen percentiles, represents a better alternative, can on the one hand be justified by its special characteristics. Thus, for example, extremely highly cited papers can hardly distort percentile-based indicators. But since every standardised indicator has its specific advantages and disadvantages, there is no indicator which is entirely without drawbacks. In order to check whether a specific indicator actually measures what it claims to measure (here: the impact of papers as a partial aspect of quality - independent of the time and subject factors), it is usual in psychometry to check the concurrent validity of the indicator. Here it is a question of how far an indicator correlates with an external criterion. Since the most important procedure for the assessment of research is the peer review procedure, the current study calculates the relation between the judgement of peers and a series of standardised indicators. Unlike with observed citation counts, we can assume that the judgement of peers is dependent neither on the subject category nor on the publication year. So the more strongly an indicator correlates with the judgement of peers, the better it appears to be suited for the measurement of impact.

In the current study, a series of cited-side and citing-side indicators are taken into account in testing their validity. Besides the normalized indicators observed citation counts have also been considered for comparison. As the results of the evaluations show, the validity of the indicators seems to be very similar - especially concerning papers assessed as "good" or "very good" by faculty members. Only for papers assessed as "exceptional" by members do greater differences appear between the indicators. With these papers, observed citation counts and the SNCSs seem to have an advantage over the other indicators for impact measurement. However, the results of this study suggest that overall, all the indicators involved here measure the normalized impact similarly - if we enlist the judgement of peers as an external criterion for the validity of the indicators.

The results of the current study could be interpreted to indicate that the method of normalization (with the indicators used in this study) has only a slight influence on the validity of the indicators. Although the F1000 papers belong to 627 different subject categories and subject category combinations, respectively, with different mean citation rates (see Table 2), the results also point out that observed citation counts perform similarly to the normalized indicators. Especially, this latter result points to some important limitations of the study:

(1) The F1000 papers are all connected to biomedical research and therefore do not reflect the true diversity of science, which normalization methods are designed to overcome. Although empirical studies including a broad range of disciplines are desired, corresponding datasets (with judgements of peers for single papers) are however not available – the F1000 dataset is a unique exception. (2) Reviewers' ratings in F1000 are given on a rather coarse scale, with just three possible levels ('good', 'very good', and 'exceptional'). A finer scale would allow a better evaluation of the indicators. (3) Using expert judgments, it is generally difficult to argue for the superiority of a normalization method – given the low reliability of expert judgments among themselves (Bornmann, 2011). A publication that is considered 'exceptional' by one reviewer may be considered just 'good' by another (Bornmann, in press). Yet another reviewer may not even consider the publication to be worth giving a recommendation in F1000. (4) The good result for the observed citation counts in comparison with the normalized indicators might be due to the fact that the judgements of the F1000 members (in the post-publication peer review process) are not only influenced by their reading of a specific paper but also by available impact data for this paper (citation counts for a short citation window and the JIF of the publishing journal).

The fact that the analysis shows no substantial differences between the different indicators can be interpreted in two ways: One interpretation is that indeed it doesn't make much difference which indicator is used. The good result for the citation indicator in this study could even mean that normalization doesn't improve the correlation of citation-based indicators and peer judgments, at least not for the highest quality publications. Perhaps artificial and questionable elements included in normalization procedures (e.g., the use of WoS subject categories) distort the outcomes of these procedures and – in some cases – cause normalized indicators to be inferior to observed citations. Given the limitations of the F1000 dataset, another interpretation seems to be also possible: the accuracy and reliability of the dataset is insufficient to distinguish between the different indicators and to make accurate comparisons between different normalized citation impact indicators. Thus, for future studies comparing judgements of experts and bibliometric indicators, datasets are necessary which cover a broad range of different disciplines.

Besides the method of normalization there are also other problems of the normalization of impact which need to be solved in future studies. With the cited-side indicators we have, for example, the problem of the journal sets, which may often be used for the field delineation of papers, but which reach their limits with small fields or multi-disciplinary journals (Bornmann, Mutz, Neuhaus, & Daniel, 2008; Ruiz-Castillo & Waltman, 2014). Another problem is the level of field delineation: for every level of field delineation there is a sub-field level, each of which generally exhibits a different citation rate. So far it has not been clarified on which level normalization should actually be performed (Adams, Gurney, & Jackson, 2008; Zitt, Ramanana-Rahary, & Bassecoulard, 2005). Finally we have the problem of the other factors which - besides the subject area and the publication year - have an influence on citation impact (independent of their quality). Future studies should investigate whether the involvement of these (and possibly other) factors is actually necessary.

# Acknowledgements

We would like to thank Ros Dignon and Iain Hrynaszkiewicz from F1000 for providing us with the F1000Prime data set.

The bibliometric data used in this paper are from an in-house database developed and maintained by the Max Planck Digital Library (MPDL, Munich) and derived from the Science Citation Index Expanded (SCI-E), Social Sciences Citation Index (SSCI), Arts and Humanities Citation Index (AHCI) prepared by Thomson Reuters (Philadelphia, Pennsylvania, USA).

# References

Adams, J., Gurney, K., & Jackson, L. (2008). Calibrating the zoom — a test of Zitt's hypothesis. *Scientometrics, 75*(1), 81-95. doi: 10.1007/s11192-007-1832-7.

Bornmann, L. (2011). Scientific peer review. *Annual Review of Information Science and Technology, 45*, 199-245.

Bornmann, L. (in press). Inter-rater reliability and convergent validity of F1000Prime peer review. *Journal of the Association for Information Science and Technology*.

Bornmann, L., de Moya Anegón, F., & Leydesdorff, L. (2012). The new Excellence Indicator in the World Report of the SCImago Institutions Rankings 2011. *Journal of Informetrics, 6*(2), 333-335. doi: 10.1016/j.joi.2011.11.006.

Bornmann, L., & Leydesdorff, L. (2013). The validation of (advanced) bibliometric indicators through peer assessments: A comparative study using data from InCites and F1000. *Journal of Informetrics, 7*(2), 286-291. doi: 10.1016/j.joi.2012.12.003.

Bornmann, L., Leydesdorff, L., & Mutz, R. (2013). The use of percentiles and percentile rank classes in the analysis of bibliometric data: opportunities and limits. *Journal of Informetrics, 7*(1), 158-165.

Bornmann, L., Leydesdorff, L., & Wang, J. (2013). Which percentile-based approach should be preferred for calculating normalized citation impact values? An empirical comparison of five approaches including a newly developed citation-rank approach (P100). *Journal of Informetrics, 7*(4), 933-944. doi: 10.1016/j.joi.2013.09.003.

Bornmann, L., & Marx, W. (in press). The wisdom of citing scientists. *Journal of the American Society of Information Science and Technology*.

Bornmann, L., & Mutz, R. (in press). From P100 to P100': conception and improvement of a new citation-rank approach in bibliometrics. *Journal of the American Society of Information Science and Technology*.

Bornmann, L., Mutz, R., Neuhaus, C., & Daniel, H.-D. (2008). Use of citation counts for research evaluation: standards of good practice for analyzing bibliometric data and presenting and interpreting results. *Ethics in Science and Environmental Politics, 8*, 93-102. doi: 10.3354/esep00084.

Bornmann, L., & Williams, R. (2013). How to calculate the practical significance of citation impact differences? An empirical example from evaluative institutional bibliometrics using adjusted predictions and marginal effects. *Journal of Informetrics, 7*(2), 562-574. doi: 10.1016/j.joi.2013.02.005.

Cohen, J. (1988). *Statistical power analysis for the behavioral sciences* (2nd ed.). Hillsdale, NJ, USA: Lawrence Erlbaum Associates, Publishers.

Colledge, L. (2014). *Snowball Metrics Recipe Book*. Amsterdam, the Netherlands: Snowball Metrics program partners.

Council of Canadian Academies. (2012). *Informing research choices: indicators and judgment: the expert panel on science performance and research funding.* . Ottawa, Canada: Council of Canadian Academies.

Egghe, L. (2005). *Power laws in the information production process: Lotkaian informetrics*. Kidlington, UK: Elsevier Academic Press.

F1000. (2012). What is F1000? Retrieved October 25, from http://f1000.com/about/whatis

Garfield, E. (1979). *Citation indexing - its theory and application in science, technology, and humanities*. New York, NY, USA: John Wiley & Sons, Ltd.

Hazen, A. (1914). Storage to be provided in impounding reservoirs for municipal water supply. *Transactions of American Society of Civil Engineers, 77*, 1539-1640.

Hirsch, J. E. (2005). An index to quantify an individual's scientific research output. *Proceedings of the National Academy of Sciences of the United States of America, 102*(46), 16569-16572. doi: 10.1073/pnas.0507655102.

Hosmer, D. W., & Lemeshow, S. (2000). *Applied logistic regression* (2. ed.). Chichester, UK: John Wiley & Sons, Inc.

Kreiman, G., & Maunsell, J. H. R. (2011). Nine criteria for a measure of scientific output. *Frontiers in Computational Neuroscience, 5*. doi: 10.3389/fncom.2011.00048.

Leydesdorff, L., & Bornmann, L. (2011). How fractional counting of citations affects the Impact Factor: normalization in terms of differences in citation potentials among fields of science. *Journal of the American Society for Information Science and Technology, 62*(2), 217-229. doi: 10.1002/asi.21450.

Leydesdorff, L., Radicchi, F., Bornmann, L., Castellano, C., & de Nooy, W. (in press). Field-normalized impact factors: a comparison of rescaling versus fractionally counted IFs. *Journal of the American Society for Information Science and Technology*.

Marx, W., & Bornmann, L. (in press). On the causes of subject-specific citation rates in Web of Science. *Scientometrics*.

Organisation for Economic Co-operation and Development. (2007). *Revised field of science and technology (FOS) classification in the Frascati manual*. Paris, France: Working Party of National Experts on Science and Technology Indicators, Organisation for Economic Co-operation and Development (OECD).

Ruiz-Castillo, J., & Waltman, L. (2014). Field-normalized citation impact indicators using algorithmically constructed classification systems of science. Retrieved September 11, 2014, from http://e-archivo.uc3m.es/handle/10016/18385

Schreiber, M. (2014). Examples for counterintuitive behavior of the new citation-rank indicator P100 for bibliometric evaluations. *Journal of Informetrics, 8*(3), 738-748. doi: http://dx.doi.org/10.1016/j.joi.2014.06.007.

Schreiber, M. (in press). Is the new citation-rank approach P100' in bibliometrics really new? *Journal of Informetrics*.

Schubert, A., & Braun, T. (1986). Relative indicators and relational charts for comparative assessment of publication output and citation impact. *Scientometrics, 9*(5-6), 281-291.

SCImago Reseach Group. (2013). *SIR World Report 2013*. Granada, Spain: University of Granada.

Seglen, P. O. (1992). The skewness of science. *Journal of the American Society for Information Science, 43*(9), 628-638.

Sheskin, D. (2007). *Handbook of parametric and nonparametric statistical procedures* (4th ed.). Boca Raton, FL, USA: Chapman & Hall/CRC.

StataCorp. (2013). *Stata statistical software: release 13*. College Station, TX, USA: Stata Corporation.

Waltman, L., Calero-Medina, C., Kosten, J., Noyons, E. C. M., Tijssen, R. J. W., van Eck, N. J., . . . Wouters, P. (2012). The Leiden Ranking 2011/2012: data collection, indicators, and interpretation. *Journal of the American Society for Information Science and Technology, 63*(12), 2419-2432.

Waltman, L., & Costas, R. (2014). F1000 recommendations as a potential new data source for research evaluation: a comparison with citations. *Journal of the Association for Information Science and Technology, 65*(3), 433-445.

Waltman, L., & van Eck, N. J. (2013a). Source normalized indicators of citation impact: an overview of different approaches and an empirical comparison. *Scientometrics, 96*(3), 699-716. doi: 10.1007/s11192-012-0913-4.

Waltman, L., & van Eck, N. J. (2013b). *A systematic empirical comparison of different approaches for normalizing citation impact indicators.* Paper presented at the Proceedings of ISSI 2013 - 14th International Society of Scientometrics and Informetrics Conference.

Waltman, L., van Eck, N. J., van Leeuwen, T. N., Visser, M. S., & van Raan, A. F. J. (2011). Towards a new crown indicator: some theoretical considerations. *Journal of Informetrics, 5*(1), 37-47. doi: 10.1016/j.joi.2010.08.001.

Wets, K., Weedon, D., & Velterop, J. (2003). Post-publication filtering and evaluation: Faculty of 1000. *Learned Publishing, 16*(4), 249-258.

Williams, R. (2012). Using the margins command to estimate and interpret adjusted predictions and marginal effects. *The Stata Journal, 12*(2), 308-331.

Williams, R., & Bornmann, L. (2014). The substantive and practical significance of citation impact differences between institutions: guidelines for the analysis of percentiles using effect sizes and confidence intervals. In Y. Ding, R. Rousseau & D. Wolfram (Eds.), *Measuring scholarly impact: methods and practice* (pp. 259-281). Heidelberg, Germany: Springer.

Wouters, P., & Costas, R. (2012). *Users, narcissism and control – tracking the impact of scholarly publications in the 21st century*. Utrecht, The Netherlands: SURFfoundation.

Zitt, M., Ramanana-Rahary, S., & Bassecoulard, E. (2005). Relativity of citation performance and excellence measures: From cross-field to cross-scale effects of field-normalisation. *Scientometrics, 63*(2), 373-401. doi: DOI 10.1007/s11192-005-0218-y.

Zitt, M., & Small, H. (2008). Modifying the journal impact factor by fractional citation weighting: The audience factor. *Journal of the American Society for Information Science and Technology, 59*(11), 1856-1860. doi: 10.1002/asi.20880.